# Using Molecular Solids as Scaled-up Qubits

Malcolm P. Roberts

# Table of Contents

# I. Introduction

The unusual properties of quantum qubits are best evidenced in environments which protect these from undesirable motions and/or interactions. These systems are usually built by trapping ions, atoms, molecules in a field or in a solid substrate to observe the desired I0> → I1> transition and perform reliable measurements on them. Pair-wise interactions between qubits have been achieved mainly by focusing on transitions in the microwave frequency range using several techniques such as silicon spin, superconducting loops, ion traps, diamond vacancies and topological qubits. The initial goal is to incorporate a much larger number of qubits, at least 50, to be able to solve some problems classical applications cannot solve, to then aggregate hundreds and eventually thousands of qubits to do meaningful work.

There is considerable work being done in scaling-up the number of qubits whilst trying to achieve controlled interactions amongst them. Aggregating a greater number of qubits can generate interactions with their environment which can undermine and even eliminate the required properties of the desired I0> → I1> transition. Severe ultra-cold conditions (fractions of a degree Kelvin), ultra-high vacuum and external fields are some of the tools used to try to confer a collection of qubits, which in essence is a collection of particles in a gaseous state, lesser degrees of freedom to try to cancel unwanted interactions. Aggregating a large number of qubits requires, amongst other things, imparting the collection of particles some degree of spatial order similar to that found in solids, or alternatively, placing individual particles directly on a solid substrate at specific locations.

The approach proposed here is opposite to the ones described above. The starting point is to use as qubits molecular solids that exhibit at very low temperatures some of the added degrees of freedom typically associated with a molecular gas. Said solids would contain an enormous number qubits from their inception and a multiplicity of transitions of interest for each qubit such as vibrations, rotations, overtone, bending and phonon modes which can be accessed at various frequencies using a multiplicity of well established experimental techniques. The rationale is to generate a "matrix" of qubits that traps the more mobile qubits. Amongst other things, the matrix provides phonon modes, spatial containment of the qubits, a means to regulate their inter-qubit separation and symmetry/anisotropy to quantum motions such as vibrations and rotations.

The simplest case to consider is an ionic salt, which can be conceived as the cation being trapped by the anion. An example is to "trap" atomic ions such as potassium, $K^+$, by the matrix generated by its anion in one of its salts such as potassium tetraphenylborate, $KB(C_6H_5)$, hereinafter KTPB . Similarly, this approach can also be used to trap highly symmetric molecules such as spherical and symmetric tops in matrices to be able to access a much larger number of I0> → I1> transitions of interest at different frequencies.

The ammonium ion, $NH_4^+$, or any of the isotopically substituted versions such as the $NH_3D^+$ ion can be trapped in matrices formed by anions such as perchlorate ($ClO_4^-$), tetrafluoroborate ($BF_4^-$) or tetraphenylborate ($B(C_6H_5)^-$). For neutral molecules such as methane, $CH_4$, or any of its isotopically substituted versions such as $CH_3D$, these matrices are solids such as Xe or pure solid $CH_4$ in its phase II. Well defined quantum motions of the matrix, such as the ones observed for the $B(C_6H_5)^-$anion at temperatures ranging from very low to room temperature, are also of interest and make the anion a matrix of qubits.

There has been a long standing interest in the theoretical and experimental understanding of molecular motion in condensed phases. However, the existence of molecular solids with components that exhibit large amplitude motions (of greater amplitude than a libration about a fixed position) at low temperatures seemed unlikely. The rotational motion in solids of molecular groups such as $-CH_3$ in a variety of alkanes [1], the vibration-rotation motion of dilute $CH_4$ in a variety of rotationally inert matrices such as nitrogen [2], xenon, krypton and argon [3]; that of $NH_4^+$ ions in its salts [4-6] have been studied by examination of C-H, C-D or N-H and N-D stretches and bends.

The possibility of ammonium ions ($NH_4^+$) freely rotating in some phases of its salts at low temperatures was first postulated by Pauling in 1930 [7]. However, the first evidence of nearly free rotation at low temperatures was not found until the theoretical and experimental study of the dynamics of $NH_4^+$ in ammonium tetraphenylborate, $NH_4B(C_6H_5)_4$, hereinafter ATPB, using infrared (IR) spectroscopy [8]. Subsequently, complementary quantum motions of the cation and anion in ATPB were studied in a series of papers quasielastic (QNS) and inelastic neutron scattering (INS), Raman and IR spectroscopies [9-12]. These papers describe in detail the experimental setups and sample preparation as well as the development of the theoretical model used to explain and assign the experimental results. A brief description of the relevant results of each paper is provided below.

The unexpected multiplicity of bands observed in the IR N-H stretching and H-N-H bending regions of the $NH_4^+$ ion in ATPB at temperatures as low as 2.6 K prompted a careful examination of the IR spectra of ATPB, and additionally that of ion in a matrix with deuterated phenyl rings, $NH_4B(C_6D_5)_4$, (ATPB-$d_{20}$), the $NH_4^+$ ion dilute in the nearly isomorphous KTPB, and the $NH_3D^+$ ion dilute in ATPB [8]. The frequency of the N-D stretch is characterized by empirical correlations as being in the range expected for $NH_3D^+$ ions undergoing free rotation [4]. The N-H stretching region shows a multiple peak structure qualitatively similar to that expected of a free rotor. The structure on the bands persists on deuteration of the phenyl groups showing that it is indeed due to the $NH_4^+$ ions. Moreover, the bands persist on dilution of $NH_4^+$ ions in KTPB showing that it is due to individual $NH_4^+$ ions [8].

ATPB belongs to a set of ammonium salts with high coordination number [4] and for these salts, the crystal structure is not obviously determined by the requirements of hydrogen bonding, assumption which had been previously used to assign the observed spectra [13,14]. The barrier to rotation of the $NH_4^+$ ion, in this case about their $C_3$ axes, has no simple relation to the hydrogen bond strength alone given that the ammonium ion sits in a position determined by the full three-dimensional angular-dependent potential of both long- and short-range forces. The appropriate symmetry group for a guest molecule that has the symmetry of the point group $\overline{M}$ (for molecule) occupying a site in a host crystal that has the symmetry of the point group S (for site) in the absence of the guest molecule depends on how much freedom the molecule has for rotation in the crystal [8].

To study of the dynamics in these crystals the "free rotor group", denoted $G = (S,\overline{M})$, was considered. $G$ is the combination of the permutations plus the permutation inversions in both $S$ and $\overline{M}$, respectively [8]. The principal free rotation group of interest for the $NH_4^+$ ion in ATPB is the one conformed for a tetrahedral molecule in a tetrahedral site, i.e., $G = (T_d,\overline{T}_d)$. For the study of the IR allowed N-H stretching motion both the energy level diagram and the intensities of the bands at low temperatures for $G = (T_d,\overline{T}_d)$ were calculated and matched experimental results. These are described in more detail in the next section. It is noteworthy that this same theoretical framework developed to study ATPB allowed for the calculation of frequency and intensity of the vibration-rotation bands for the C-H stretching region of $CH_4$ dilute in a Xe matrix and reassign the IR stretching spectra [15]. This is possible given that the C-H vibration-orientation transitions of $CH_4$ dilute in a Xe matrix are similar in nature to the N-H ones in ATPB.

The second and third papers consider the quasielastic neutron scattering (QNS) and inelastic neutron scattering (INS) data for ATPB and ATPB-$d_{20}$, respectively [9, 10]. The QNS data show that for temperatures of 120 K or higher, all the $NH_4^+$ ions reorient about a $C_2$ axis with an associated activation energy of about 240 $cm^{-1}$. However, for temperatures between 80 and 40 K, the calculated activation energy decreases to half its high temperature value [9]. The low temperature INS spectrum shows the existence of two inequivalent $NH_4^+$ ions, present in about equal percentages. The rotational tunneling spectra observed in the μeV region from 1.5 to 40 K characterizes one kind of $NH_4^+$ ion as librating is a cubic site associated with a barrier of about 280 $cm^{-1}$, a value very close to the high temperature activation energy calculated from the QNS data. The INS data in the meV region shows two sharp bands that can be interpreted in two ways: either the 0-2 and 0-3 rotation transitions of the nearly free rotating second kind of $NH_4^+$ ion; or on the other hand, as arising from two ions that librate, one in a well associated with the tunneling spectrum and one of a second kind of ion librating in a deeper well [10].

In a fourth paper the Raman spectra of both the N-H stretches and low frequency modes in which both the cation and anion participate are presented and assigned [11]. Several aspects of these spectra turn out to be of interest: Firstly, many bands can be assigned as arising from the nearly free rotational (orientational) motion of the $NH_4^+$ ion, and in particular, the predicted orientational transitions are observed directly in the low frequency region. This resolves the ambiguity in the interpretation of the INS meV data. Secondly, the presence of certain bands shows that the $NH_4^+$ ion is affected by $D_{2d}$ perturbations arising from interactions with the crystallographic site. Together, these phenomena help contemplate the description of this unique crystal containing a "nearly free rotor." Thirdly, the combination of the low frequency Raman spectra of KTPB, ATPB and ATPB-$d_{20}$ and the INS results allows to assign the low frequency anion bands.

A fifth paper considers the effect of the $D_{2d}$ perturbations [12]. The small bands at higher frequency in the N-H stretching region are not fitted by the theoretical model and were assigned to "lattice –sum bands" [8]. When ATPB is recrystallized in a controlled manner, bands in the high frequency region change in shape and increase in intensity and number, as observed in the IR and Raman N-H stretching spectra of the $NH_4^+$ ion in ATPB, ATPB-$d_{20,}$ and dilute in KTPB. The N-D stretching of the $NH_3D^+$ ion dilute in ATPB shows even more complex recrystallization effects. These high frequency bands which are superimposed to the weaker lattice-sum bands are explained as arising from a second N-H and N-D stretching vibration which is allowed when molecules with either $T_d$ ($NH_4^+$) or $C_{3v}$ ($NH_3D^+$) symmetry are in a lattice site which has a higher $D_{2d}$ character. The character tables for ($D_{2d}$,$\overline{T}_d$) and ($D_{2d}$,$\bar{C}_{3v}$) were developed to complete the assignment. Additionally, the spectra of $NH_4ClO_4$ and of solid $CH_4$ II are also examined and explained following a similar methodology [12].

The molecules in the solids described in the five papers summarized above represent the first examples of qubits being proposed using this approach, especially because they contain molecules which exhibit rare large amplitude motions at low temperatures in a controlled environment.

Even before considering the specific detail of the transitions of interest of the molecules presented as candidate qubits, advantages to this approach can already be identified:

(i) Neither external fields nor ultra-high vacuum conditions are required for trapping purposes;

(ii) There are a multiplicity of I0> → I1> transitions of interest which are provided by each and every one of the qubits present in the solid; and

(iii) In addition of granting qubits spatial order within a given crystal, solids can also grant qubits distinct locations in space when forming an array of crystals.

These generic advantages open several possibilities, such as being able to differentiate otherwise identical transitions in space, to develop novel and diverse experimental hardware and control gates in more benign and stable experimental conditions, to enable parallel computing and split photon excitations to generate entanglement of molecular transitions at a distance, etc.

The scaled-up qubits being proposed, hereinafter also called aggregated qubits, perform several of the I0> → I1> transition of interest. Some of the specific properties of the proposed qubits are identified and examined in the next section.

## II. Proposed Qubits

The first scaled-up qubits being proposed comprise both the components of the matrix and the ensemble of atoms and/or molecules trapped by the matrix. The specific examples which are discussed here are the components of ATPB, ATPB-$d_{20}$, KTPB and the $NH_4^+$ ion dilute in KTPB.

Additionally, qubit candidates which are proposed but left to be examined in further detail in future studies comprise: (i) The isotopic variants of the $NH_4^+$ ion, such as the $NH_3D^+$ ion, dilute in any of the solids mentioned above; (ii) the $K^+$ ion; and (iii) $CH_4$ either in its solid phase II or dilute in a matrix such as Xe.

In Section A below some of the general properties of these type of qubits are identified. In Section B some of the multiple transitions of interest are examined in detail.

## A. General Properties of Candidate Qubits

### 1. Avoidance of working in ultra-cold conditions

As described in the introduction, the quantum motions of these solids are well characterized in experiments performed at temperatures of several degrees Kelvin, starting as low as 1.5 K and in the case of the lattice modes, all the way up to room temperature. It is therefore a salient advantage of this approach to be able to work with stable quantum transitions which do not require ultra-cold conditions to be observed and measured.

### 2. Larger array of experimental techniques are applicable

The transitions of interest have been studied both theoretically and experimentally and can be accessed at a variety of frequencies within the 10 to 3300 $cm^{-1}$ range using well developed experimental techniques such as IR, Raman, NMR, neutron scattering, microwave, etc.

### 3. Multiplicity of l0> → l1> transitions

All of the proposed candidate qubits have multiple I0> → I1> transition of interest. The case of selecting only a single specific transition of interest would correspond to maintaining a one-to-one relationship between the qubit and the transition of interest, which is usually the case in most of the ongoing research.

The current approach has a much larger array of possibilities given that there are at least two different qubits present in the solid and additionally, each qubit has multiple and distinct transitions of interest. This opens up the potential to combine and control diverse I0> → I1> transitions of interest in ways which are not possible at present.

As an example of number and variety of transitions, the spherical top candidate in ATPB, the $NH_4^+$ ion, has many types of transitions [8] such as vibration, overtone, bending, rotation and couplings of some of these motions to lattice modes which could be of interest in diverse applications. Likewise, the $B(C_6H_5)_4^-$ ion has a variety of phonon/normal modes that are well characterized from low to room temperature and are discussed in section B below.

Moreover, some of the transitions, such as the $\nu_3$ N-H stretching I0>→I1> transitions of the $NH_4^+$ ion in ATPB, ATPB-$d_{20}$ and dilute in KTPB, can be conceived as several individual and distinct I0>→I1> transitions in isolation, or alternatively, as a multi-level I0>→I1>+I1'>+I1">+…vibration-orientation transition. In principle the $\nu_3$ N-H stretching I0>→I1> transition of this spherical top should be a single band at low temperatures in the librator limit, but instead shows a multiplicity of bands which arise from the coupling of the stretch to the rotations about the $C_3$ axes of the molecule given its proximity to nearly free rotation in an environment of high symmetry provided by the matrix formed by the anions. These vibration-orientation transitions are examined in detail in section B below.

## 4. Chemical composition:

Except for the $K^+$ ion, all qubits have hydrogen atoms as components. This opens the possibility to partially or totally deuterate the qubit to alter transition multiplicity and frequencies. This effect has been exploited to some degree in the study of dynamics of ATPB [8-12]. Additionally, the presence of hydrogen atoms allows for great sensitivity to NMR pulses and control techniques and greater neutron scattering cross-sections.

## 5. Aggregation:

A fundamental outstanding quest is to reach a critical aggregate of qubits to be able to have the computing power to tackle problems that classical computers cannot solve within a reasonable timeframe or at all. The salient and common properties of the qubits presented in this study is that they are already aggregated as a very large arrays and that the quantum motions involved have been studied to great extent both theoretically and experimentally. The aggregation can be controlled in various ways, some of which are highlighted below:

### *a. Aggregation in numbers:*

Rather than talking of the transition of a single qubit, a transition within the context of these molecular solids describes the excitation of a large ensemble of qubits which respond to said excitation. In other words, all I0> → I1> transitions being considered here involve exciting a very large number of molecules for each transition frequency selected. The spectral bands in the studies discussed in the introduction and those reproduced here represent the excitation of a large aggregate of qubits ($NH_4^+$, $NH_3D^+$, $CH_4$, $B(C_6H_5)_4^-$ and $B(C_6D_5)_4^-$) in each and every frequency.

A particular advantage of ATPB is that the aggregation of the spherical top qubits, the $NH_4^+$ ion, and that of the symmetric top, such as the isotopically substituted $NH_3D^+$ions, can be controlled by diluting the ions to any desired level in KTPB and ATPB, respectively. This allows regulating the absolute number of qubits present in the crystal and, especially, the inter-qubit distance in the range from single digit micrometers for the 100% ATPB to several hundred micrometers for a dilute sample of say 1% $NH_4^+$ in KTPB, dilution which does not establish a boundary in any way. This is further described when the transitions of interest are examined in the next subsection B. The dilution of the $NH_4^+$ and/or the $NH_3D^+$ ion in KTPB can also serve as a means of preparing samples with a higher number of qubit types as it will also add the $K^+$ qubit.

This concept of dilution could be further explored for the lattice modes. Experiments were carried out with the observed lattice modes being either 100% $B(C_6H_5)_4^-$ or with samples of $B(C_6D_5)_4^-$ which gave phenyl hydrogen signals less than 0.5% of that measured for ATPB [8]. In principle, it should be possible to dilute one into the other if desired. Additionally, some of the partially deuterated species could be created to provide dilution of $B(C_6H_5)_4^-$ and/or $B(C_6D_5)_4^-$ or to generate alternative lattice transitions.

### *b. Aggregation of frequencies:*

The quantum transitions of these molecular solids result in spectra showing bandwidths that are in the order of several wavenumbers (measured as Full Width Half Maximum – FWHM) even at the lowest temperatures and highest spectral resolution. The narrowest band observed has a FWHM of 1.5 $cm^{-1}$ (some 45 GHz) corresponding to the IR N-D stretching mode of the 0.5% isotopically dilute $NH_3D^+$ molecule in ATPB at 7K. (See Fig. 2 in Ref. [8]). In general, it follows that any given I0> → I1> transition of interest is an envelope of frequencies and intensities, even if higher spectral resolution is utilized. This opens the possibility of selecting different probing frequencies within the same I0> → I1> transition when used as qubits. A narrower frequency source than band width, such as a laser, will be able to excite several transitions with different frequencies within each band envelope, thus achieving differentiation within a given transition of interest for control purposes.

The aggregation in frequencies discussed here has an added tool: rather than just excite a given frequency, tunable IR lasers could burn holes in the absorbance spectrum, therefore being able to specifically subtract frequencies from band profiles.

#### c. Aggregation of distinct crystals:

Multiple distinct crystals could be manipulated in parallel as part of a whole. The stability provided by the crystal generates a virtually endless trapping field which can be divided in subsets all with the same potential properties. For example, different transitions could be selected in distinct crystals for reasons of manipulating them with different techniques and multiple probes as a parallel array.

In addition to the above and depending on the experimental setup, split photons could also excite the same motions in different crystals thus paving the way to entanglement at a distance. Precedent for this has already been established by entangling the phonon modes of separate macroscopic diamonds at room temperature [16].

## B. Transitions of Interest

The following quantum motions will be further discussed to highlight their readiness as I0> → I1> transitions of interest: (i) the lattice modes, observed clearly using Raman spectroscopy between 66 and 160 $cm^{-1}$; (ii) the $\nu_3$ N-H stretching I0>→I1> transitions of the $NH_4^+$ ion in ATPB, ATPB-$d_{20}$ and dilute in KTPB with its main IR bands spanning from about 3200 to 3240 $cm^{-1}$; and (iii) the $NH_4^+$ ion rotations which are observed directly using Raman spectroscopy at frequencies below 60 $cm^{-1}$ and which are also present as the orientation component of the vibration-orientations in (ii) above. The C-H vibration-orientations of $CH_4$ dilute in a Xe matrix are considered transitions of interest by analogy to the N-H stretching transitions in (ii) above.

The N-D motions of the isotopically dilute $NH_3D^+$ symmetric top molecule in ATPB together with the overtone and bending modes of the $NH_4^+$ ion are also considered suitable transitions and may be preferable to the motions selected here in some experimental setups. Their spectra and analysis of their motions is highlighted in Refs. [8] and [12], and their readiness as I0> → I1> transitions of interest will be evaluated in future papers.

### 1. Lattice Modes

The low-frequency Raman and INS spectra (frequencies below 160 $cm^{-1}$) are rich with transitions which emerge from the quantum motions of both the anion and cation in ATPB and ATPB-$d_{20}$. It is noteworthy that the very large inelastic scattering cross section of the protons allows the $NH_4^+$ ion motions to practically completely dominate the INS spectrum of ATPB-$d_{20}$ [10]. In addition, the Raman spectra of the anion motions in KTPB are well resolved and are used as a reference in the assignment of the more complex ATPB and ATPB-$d_{20}$ spectra [11].

The translational symmetry of solids gives rise to the existence of low-frequency phonons. In ordered systems, either crystalline solids (as in this case) or isotropic homogeneous continua, some phonons characterized by different wavevectors are degenerate [17], which seems to be the case in these molecular systems. Six bands are Raman active under $T_d$ symmetry for the normal modes of an $XY_4$ anion. These bands are expected to split into fifteen Raman allowed bands if the $D_{2d}$ perturbation is strong enough to allow all bands to be resolved within the experimental resolution.

Fig. 1 shows the contrast between the low-frequency Raman spectrum of KTPB and ATPB at low temperatures. The spectrum of KTPB consists of six bands, in agreement with the predictions of $T_d$ symmetry. The low temperature Raman spectra of ATPB and ATPB-$d_{20}$ are more complex. They show the allowed normal modes of the anion, which in both cases include a higher number of allowed anion modes than for KTPB. These extra anion bands are interpreted as arising from $D_{2d}$ perturbations of the otherwise $T_d$ matrix.

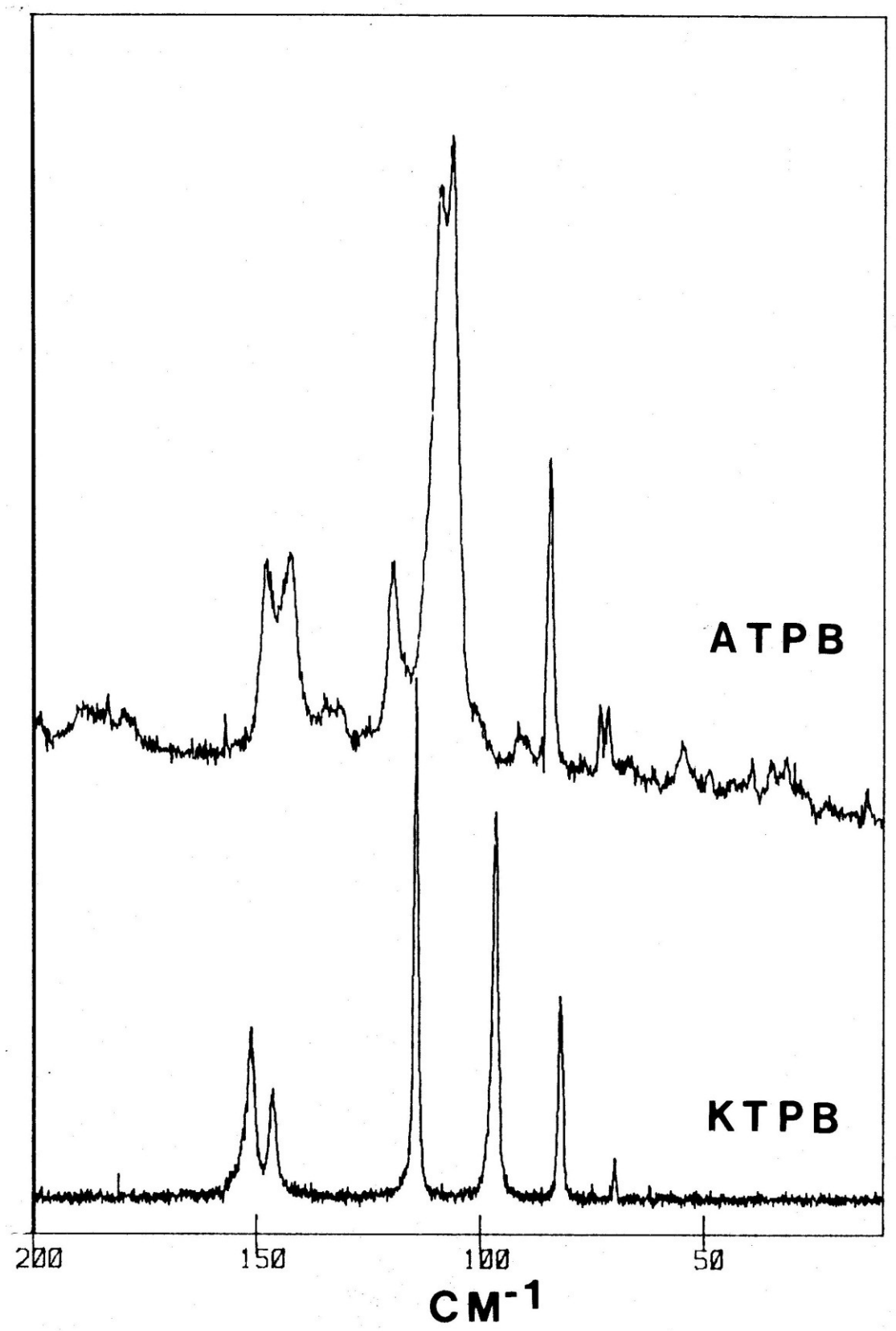

**Figure 1. The Raman low frequency spectra of ATPB at 40 K (top spectrum)and KTPB at 30 K (bottom spectrum). The spectral resolutions used are 1 and 0.7 cm$^{-1}$, respectively. (See Ref. [11])**

The low frequency Raman bands of ATPB, ATPB-$d_{20}$ and KTPB are assigned in Tables II and III of Ref. [11]. The transitions below 60 cm-1 are due to the rotational motion of the cation and are discussed in a separate section below. The anion modes are observed between 66 and 160 cm$^{-1}$ in all cases and are well defined over a very large temperature range. Fig. 2 shows the example of the spectra of ATPB-$d_{20}$ between 20 and 300K.

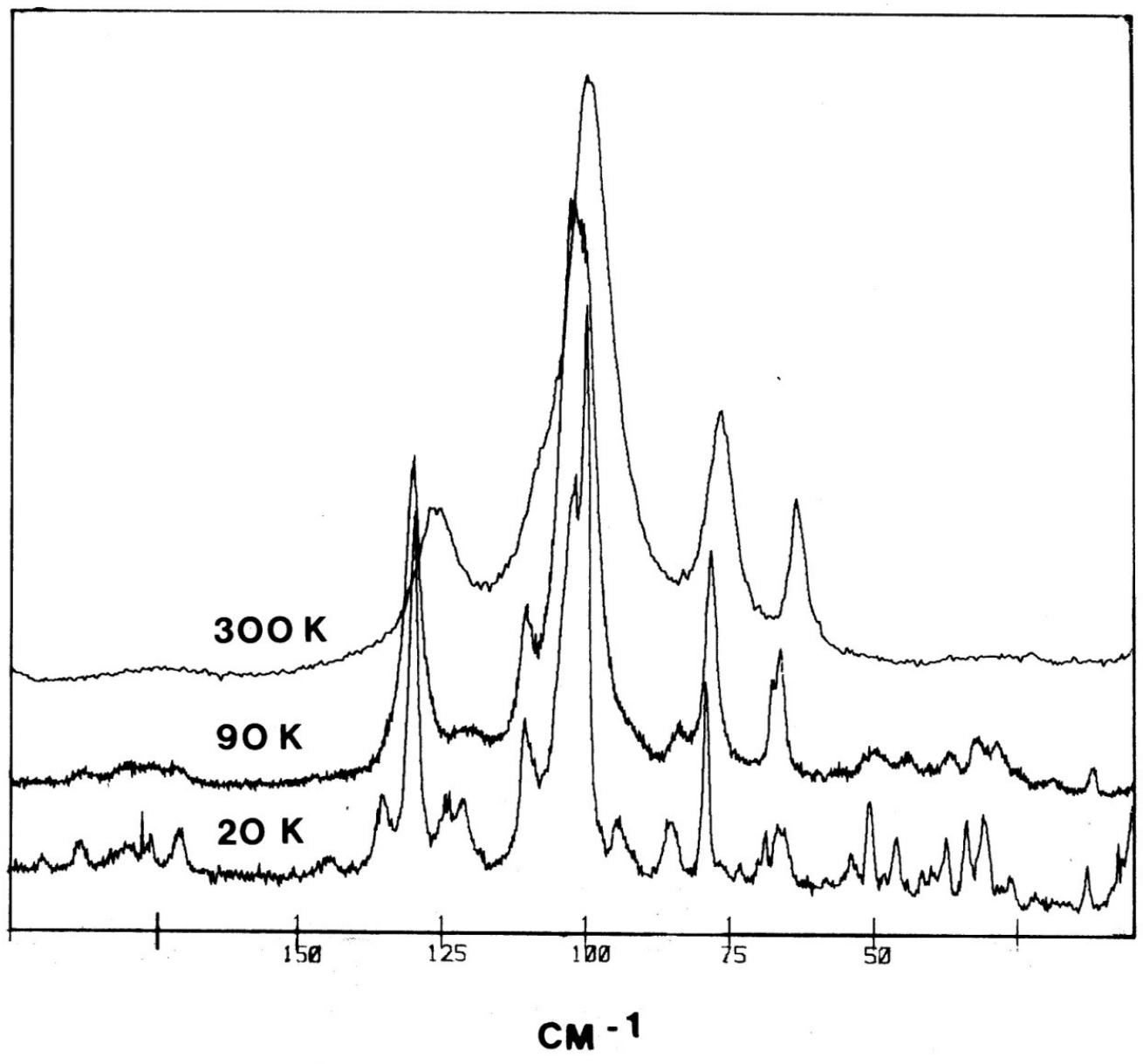

**Figure 2. The Raman low frequency spectrum of ATPB-$d_{20}$ at 20, 90, and 300 K. The sample was obtained using the controlled recrystallization method. The spectral resolution is 1cm $^{-1}$. (See Ref. [11])**

In summary, the lattice modes provide multiple lattice I0>→I1> transitions of interest which are well defined from very low temperatures to room temperature.

Additionally, there are further potential I0>→I1> transitions of interest involving lattice modes. There is evidence that some lattice motions couple with the motions of the "trapped" cation molecules to create "sum bands", for example, the vibration-orientation + lattice mode sum bands present in the crystals with higher $T_d$ anion symmetry in the IR spectra of ATPB some 66 cm$^{-1}$ higher in frequency than the $(A_1)_1 \rightarrow (L_1)_1$ vibration-orientation transition centered around 3217 cm$^{-1}$ [8, 12].

## 2. N-H vibration-orientation

These are quantum motions with dipole (IR) and quadrupole (Raman) allowed transitions. As shown in Fig. 3, the N-H stretching $\nu_3$ vibration of the $NH_4^+$ ions in ATPB shows multiple IR bands for its I0> → I1> transition even at temperatures of 2.6 K. This is in stark contrast to the single N-H stretching band which is predicted for the ion at low temperatures in the high barrier librator limit which has an effective $\bar{\bar{T}}_d$ symmetry, said point group symmetry noted by adding two bars on top.

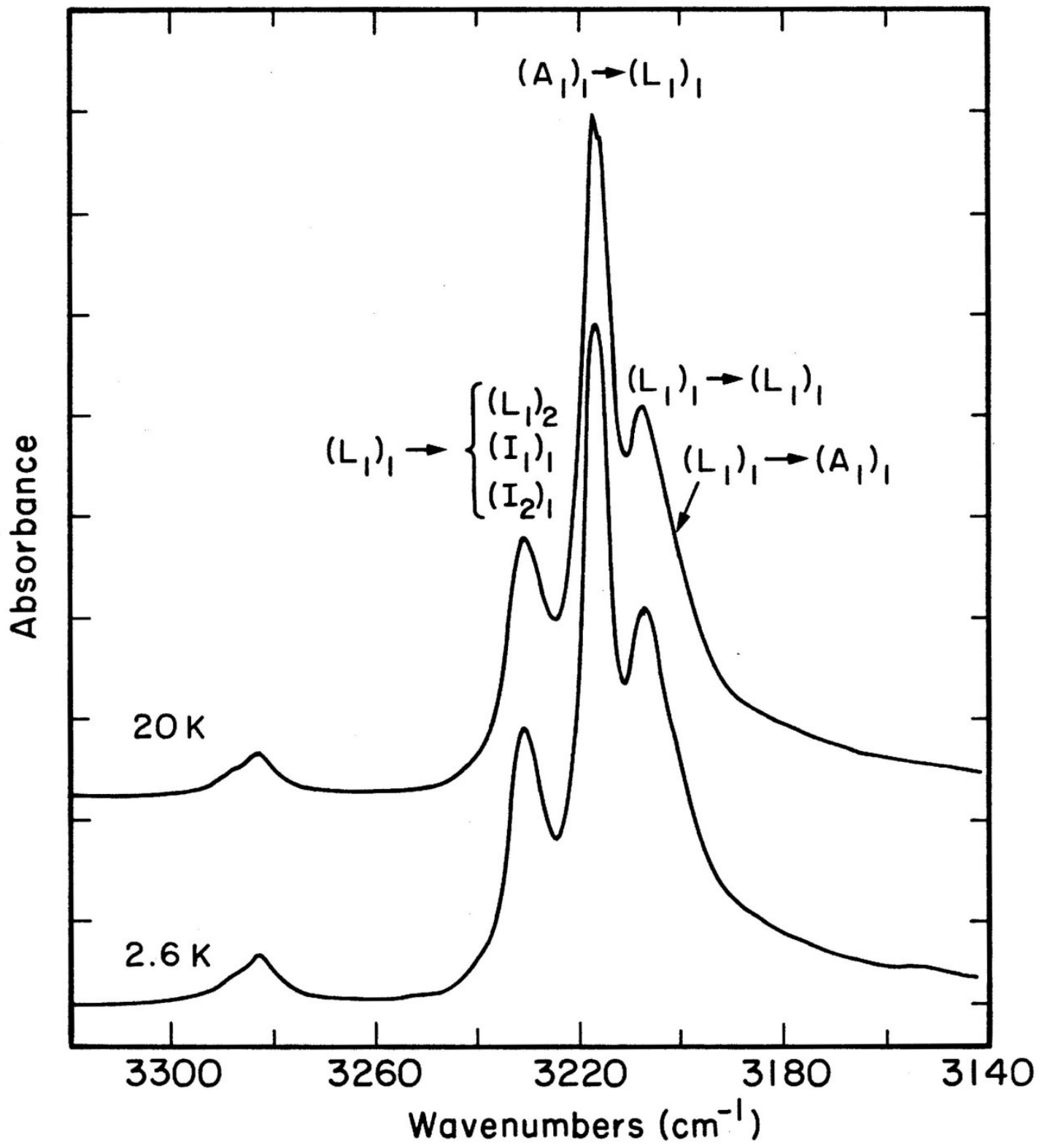

**Figure 3. The infrared spectrum of the N-H stretching region of pure ammonium tetraphenylborate (ATPB) at 2.6 and 20 K. Only the assignments of the strongest bands are indicated. The notation is that explained in Fig. 4. Note that the spectra do not change appreciably with the eight-fold increase in temperature. This is evidence for nuclear spin freezing. (See Ref. [8])**

These main bands are present in ATPB, $NH_4^+$ ions dilute to any extent in KTPB (e.g., 0,5%), ATPB-$d_{20}$ in both controlled (higher $D_{2d}$ character for the lattice symmetry) and uncontrolled (higher $T_d$ character for the lattice symmetry) crystallizations. As mentioned in the previous section, these transitions are attributed to vibration-orientations of the nearly freely rotating $NH_4^+$ ions about their $C_3$ axes and have been studied in detail. The relative intensities of the N-H stretching bands do not change appreciably when the temperature is raised from 2.6 to 20 K, as shown in Fig. 3. At higher temperatures the bands broaden but do not shift in frequency [8].

## *The Theoretical Model*

The theoretical model is described in Ref. [8] (Section III B, III C and the Appendix). The model is developed to understand the dynamics of a tetrahedral molecule in a tetrahedral external field, which form the combined group $G$ = ($T_d$,$\overline{T}_d$) in the free rotor limit which, with increased potential energy barrier, correlates to $\overline{\overline{T}}_d$ symmetry in the librator limit. The selection rules for the transitions are derived from the combined group's character table [8].

Since what is of interest is calculating the energies of the rotational levels it is desirable to work in the subgroup of the proper rotations of ($T_d$,$\overline{T}_d$), i.e., $T x \overline{T}$. The Hamiltonian operator for the libration of a spherical top:

$$H = BP^2 + \beta B V(\omega) \qquad (1)$$

where the rotational constant is defined as $B = \hbar^2/(2I)$ and is used as a unit of energy, $I$ is the moment of inertia, $P^2$ is the total angular momentum squared, and $\beta$ is a dimensionless field-strength parameter which determines the barrier to rotation. We need to solve

$$H\Psi(\omega) = \gamma B\Psi(\omega) \qquad (2)$$

where $\gamma$ is a dimensionless energy and $\omega$ are the Euler angles. The symmetry adapted spherical rotor functions $\Psi(\omega)$ were generated from symmetry adapted spherical harmonics. The Hamiltonian matrix has a diagonal kinetic energy matrix with matrix elements $BJ$ ($J$ + 1), where $J$ is the rotational angular momentum. The potential energy matrix $V(\omega)$ is the sum of potential energy matrices for each of the irreducible symmetry representations of $Tx\bar{T}$. The matrix elements are described in more detail in Ref. [8] Section III C.

The solution of Eq. (2) gives a complete set of eigenvectors and eigenvalues for each irreducible symmetry representation and for each chosen value of $\beta$. The energy level diagram in Fig. 4 is characterized by the dimensionless parameter $\beta$ =1 and shows the rotational energy levels near the free rotation limit calculated for ATPB using a rotational constant value of $NH_4^+$ of $B$= 5.799 cm$^{-1}$. The barrier to rotation about a threefold ($C_3$) $NH_4^+$ axis is calculated to be of only 14 cm$^{-1}$. The $J$ = 0 and $J$ = 1 levels are below the barrier and the $J$ = 2 levels are above it [8]. The first three levels of the ground vibrational state shown in the energy level diagram, $(A_1)_1$, $(L_1)_1$ and $(E_2)_1$ , have three different nuclear spin species, A, F and E, respectively.

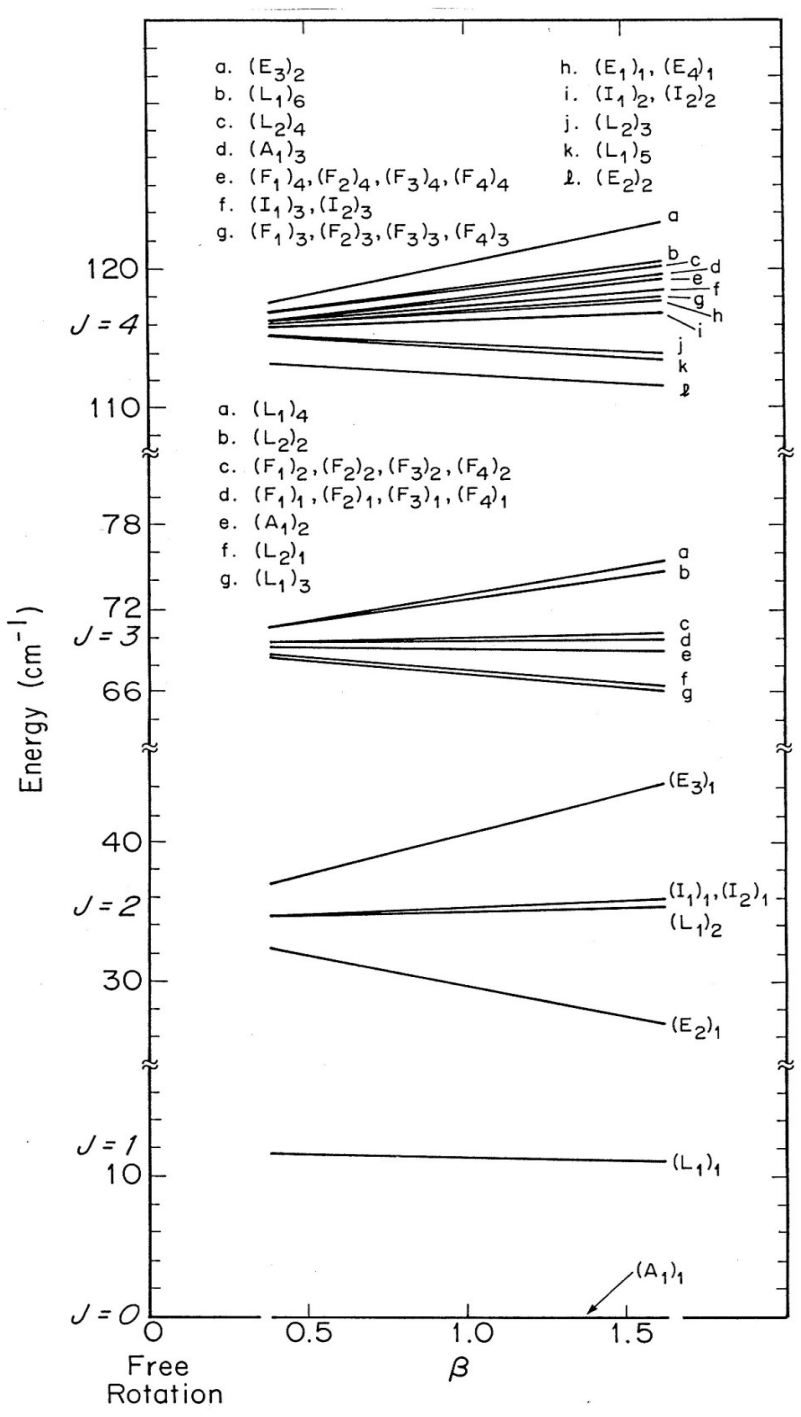

**Figure 4. Energy level diagram for a tetrahedral molecule in a tetrahedral site near $\beta$ = 1.0. The levels are labeled $(R_n)_p$, where $R_n$ is a symmetry representation of ($T_d$,$\bar{T}_d$). The subscript p is used to distinguish between energy levels with the same symmetry. The energy levels were calculated with the rotational constant $B$ = 5.799 cm-1 and the reduced potential parameter is defined in Eq. (1). (See Ref. [8])**

The intensities of the IR allowed vibration-orientation transitions at different temperatures are calculated using the methodology described in the Appendix of Ref. [8]. The calculated intensities for temperatures ranging from 7 to 50 K and its comparison to experimental intensities is given in Ref. [8] (Table IV and Fig. 6 therein). The fit of the calculated intensities and frequency spacings to the spectral bands centered around the most intense $(A_1)_1 \rightarrow (L_1)_1$ vibration-orientation transition at 3217 $cm^{-1}$ is excellent. For temperatures around 20 K and below spin freezing plays a role in the relative intensities of the main transitions, given that the first three rotational levels are populated at the lower temperatures [8].

The main transitions originating from the $(A_1)_1$, $(L_1)_1$ and $(E_2)_1$ levels in the ground V=0 vibrational state to the $J$ = 0, 1 and 2 levels of the first excited V= 1 vibrational state, together with the selection rules are summarized in Fig. 5. Taking as a reference the $(L_1)_1 \rightarrow (L_1)_1$ transition between the I0> →I1> vibrational states, the tunneling frequencies $\omega_{LA}$ and $\omega_{LE2}$ are between the $A_1$ and $L_1$ rotational state and between the $L_1$ and $E_2$ rotational state of the ground vibrational level I0>, respectively. The tunneling frequencies $\omega^*_{LA}$ and $\omega^*_{LE2}$ have the same meaning except that they are for the first excited vibrational level I1>. The values of the tunneling frequencies are not required to be identical in the ground and excited state but the experimental results show that there no evidence of them being different within the limits of the experimental resolution of the spectra, which is typically in the order of 1 $cm^{-1}$. However, measurable differences could arise when used as qubits if, for example, tunable IR lasers are used.

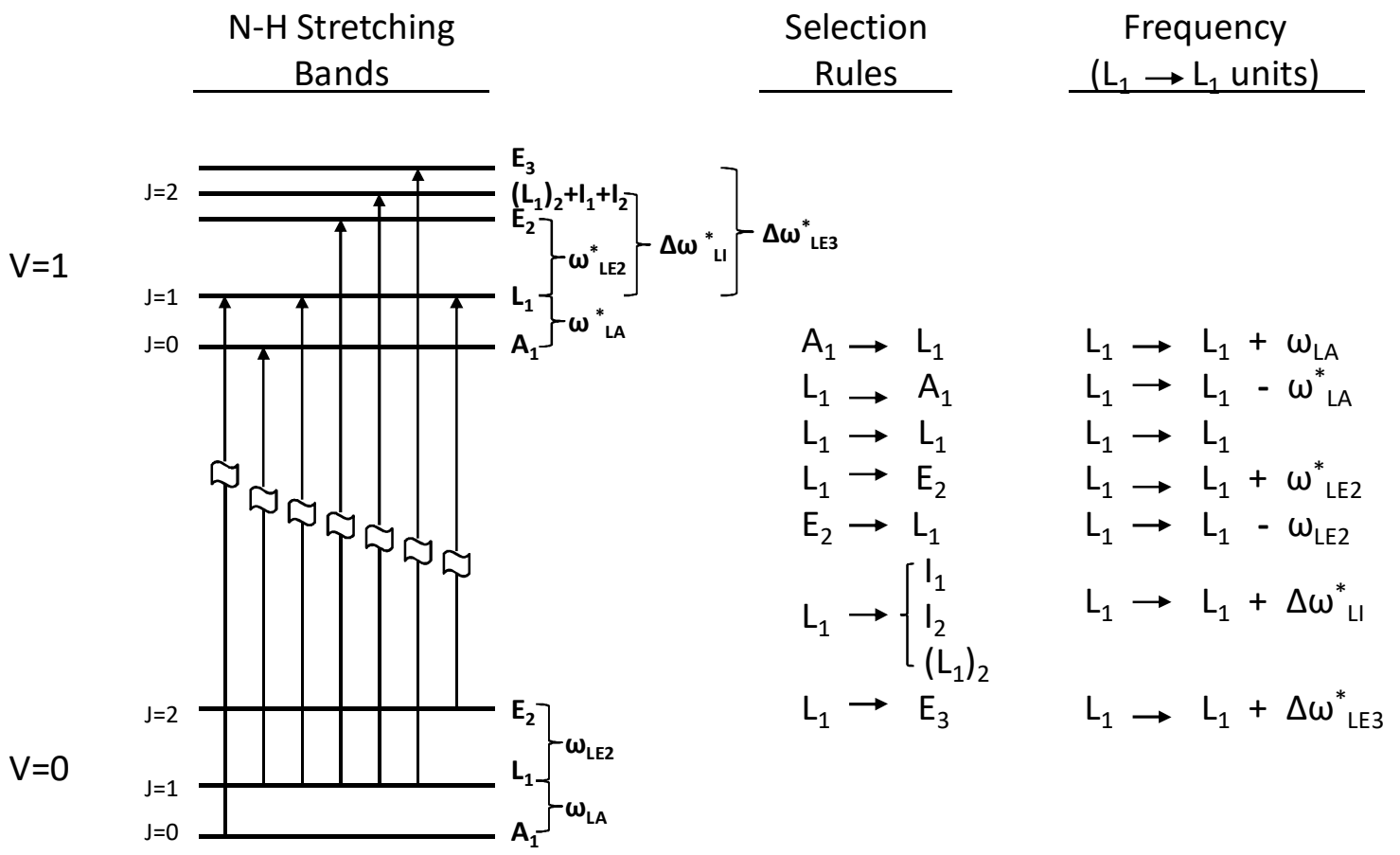

**Figure 5. The main infrared-allowed vibration-orientation transitions originating from the $(A_1)_1$, $(L_1)_1$ and $(E_2)_1$ orientation levels in the ground vibrational state (V = 0). These levels are populated at very low temperatures due to nuclear spin freezing. Only the transitions to the $J$ = 0, 1 and 2 levels of the first excited vibrational state (V = 1) are shown, together with the selection rules and relative frequencies.**

From the intensity calculation (see Ref. [8], Table IV) transitions involving the $(E_2)_1$ rotational state are predicted to be have considerably less intensity than the ones involving the $(A_1)_1$ and $(L_1)_1$ states. In addition to these five allowed vibration-orientation transitions at low temperatures there are two further transitions, both observed and calculated, the $(L_1)_1 \rightarrow (L_1)_2 + (I_1)_1 + (I_2)_1$ transition at 3230 $cm^{-}$ [1] and the $(L_1)_1$ →$(E_3)_1$ transition at 3235 $cm^{-1}$. When a higher resolution of 0.25 $cm^{-1}$ was used on a 5% NH4+ dilute in KTPB sample at 7 K, the$(L_1)_1$ →$(E_3)_1$ transition was observed as a side band of the more intense $(L_1)_1$ →$(L_1)_2 + (I_1)_1 + (I_2)_1$ transition, as shown in Fig. 6. For these transitions there are higher difference in frequency to the reference $(L_1)_1$ → $(L_1)_1$ transition, namely $\Delta\omega^*_{LI}$ and $\Delta\omega^*_{LE3}$, respectively.

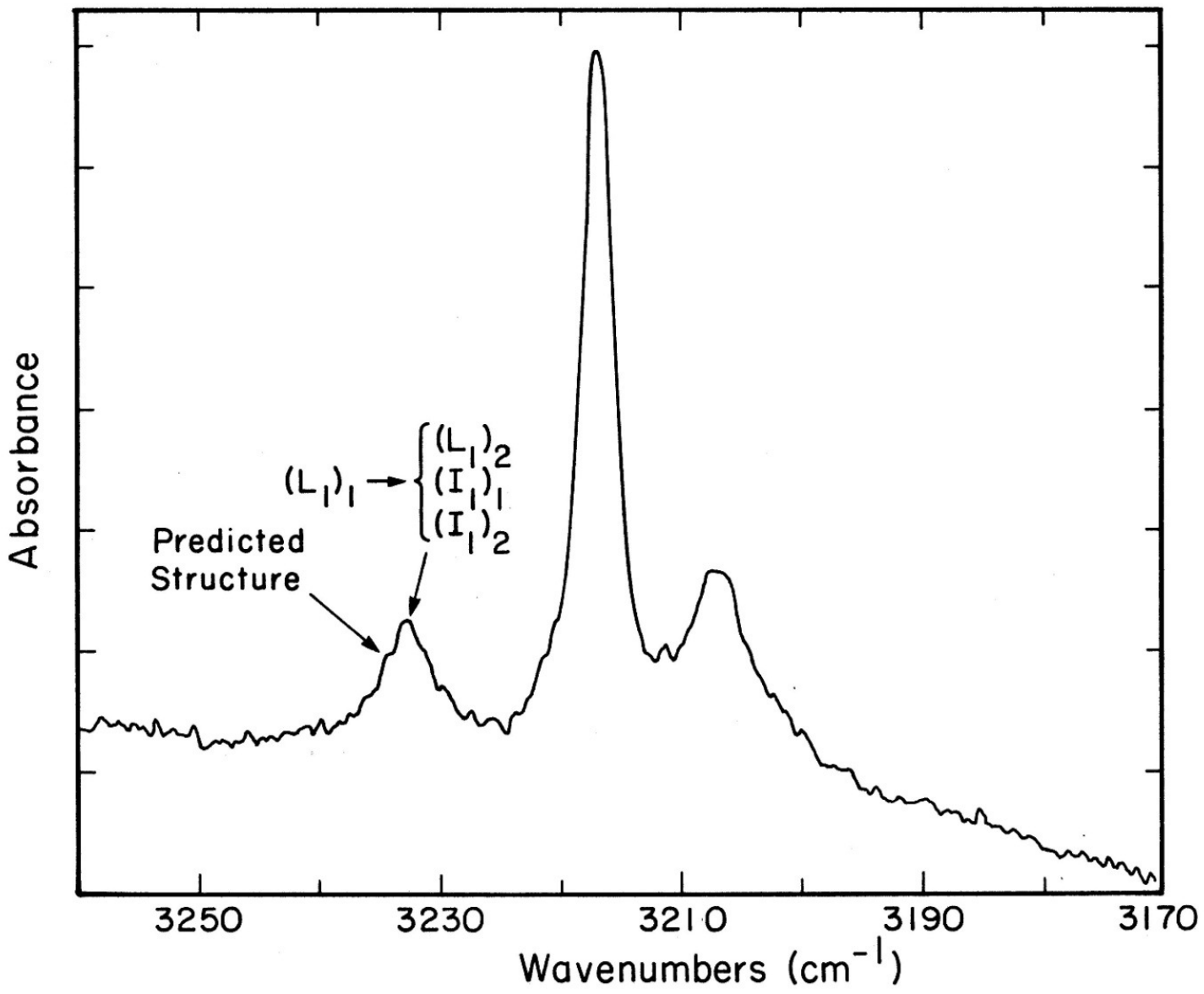

**Figure 6. A 0.25 $cm^{-1}$ spectrum of the The N-H stretching bands of 5% $NH_4^+$ dilute in KTPB at 7K. The predicted high frequency structure, which is not clearly seen in the spectra of the neat ATPB is resolved in this spectrum. The predicted structure corresponds to the $(L_1)_1$ →$(E_3)_1$ transition shown in the diagram in Fig. 5 and calculated to be at 3235 $cm^{-1}$. (See Ref. [8])**

The use of these $\nu_3$ N-H stretching vibration-orientation transitions as qubits should have several benefits in their implementation. Taking into consideration the general properties described in A above together with the theoretical model and experimental results, the following additional features are highlighted:

### *a. Multiple levels within a I0> → I1> transition*

This vibration-orientation transition can be conceived as several individual distinct vibration-orientation transitions, such as the most intense $(A_1)_1$ →$(L_1)_1$ vibration-orientation transition mentioned above, or alternatively, as a otherwise single $\nu_3$ N-H stretching I0>→I1> vibration band that has split into several I0>→I1>+I1'>+I1''>+… transitions with different orientation components, such as the collection of vibration-orientations which dominate the spectra below 16 K.

As shown in Fig. 5 and, there are several vibration-orientation transitions with distinct frequencies within the N-H stretching $\nu_3$ I0> → I1>vibration at low temperatures. In the highly symmetric environment provided by the tetraphenylborate anion and without the presence of an external field, this I0> →I1> vibration has both the ground and excited state vibrational level's degeneracy lifted by the reorientation of the molecules about their $C_3$ axes transforming what is expected to be a single stretching band at low temperatures to a multiplicity of transitions of this nearly free rotor. The resulting vibration-orientation bands at low temperatures, with separations in frequency Δω, are easily observed in the N-H stretching region, as shown in Fig. 3, whilst the orientation transitions can be observed directly in Figs. 1 and 2 as rotational bands in low frequency Raman spectra and are further discussed below in the section on $NH_4^+$ rotations.

As an example, we consider the vibration orientation I0> → I1> transitions $(L_1)_1$ →$(L_1)_1$ and the$(A_1)_1$ →$(L_1)_1$. These have been calculated and observed (within experimental resolution) at 3206 $cm^{-1}$ and 3217 $cm^{-1}$, respectively, making the difference between the I01> and I10> vibration-orientation states in the IR spectrun of $\Delta\omega = \omega_{LA} = 11\ cm^{-1}$, which are clearly distinct bands in the spectra shown in Fig. 3 and previous studies. The same procedure could be carried out with any of the other transitions listed in Fig. 5, allowing amongst other things, for the possibility of selectively exciting specific transitions without affecting others. This order of magnitude of Δω, in the hundreds of GHz, corresponding to the vibration-orientation and/or tunneling frequencies between orientational levels transitions, is a frequency difference which should open the possibility of different experimental setups to pulse, excite, measure and control with available technologies these tunneling frequencies between orientational levels of $NH_4^+$ ions in nearly free rotation and stimulate entangled states.

Additionally, the property of having multiple quantization level options in a same I0> → I1> transition opens whole new possibilities such as carrying diverse quantum information signals at once, as required in true quantum cryptography, which requires the contents of the communication to be sent in a non-classical manner. Moreover, as in cryptography, multiple quantization levels allows to conceive combined states beyond the pairwise I00> → I11> transitions, to develop both computation and logic protocols that go beyond the binary system.

### b. Spin Freezing and the Long Lifetime of the Excited States

An important characteristic of these vibration-orientations is related to lifetime of the excited states. The calculated intensities at 7 K shows that all bands should have below 5% of the intensity of the $(A_1)_1$ →$(L_1)_1$ vibration-orientation transition at 3217 $cm^{-1}$ [8]. However, as mentioned earlier, Fig. 3 shows that the main bands do not change appreciably between 2.6 and 20 K, and in those cases, are well above 5% of the relative intensity of the $(A_1)_1$ →$(L_1)_1$ transition.

Comparison between theoretical and experimental intensities of the IR vibration-orientation at temperatures around 20 K and lower show that the experimental spectra have relative intensities which correspond to higher temperature population distributions, as shown in Fig. 7. This is attributed to the slow rate of interconversion between the different nuclear spins which hinders a rapid relaxation to the ground $(A_1)_1$ state. This spin freezing partly determines the intensities of the IR bands [8].

The IR allowed $(A_1)_1$ →$(L_1)_1$ vibration-orientation transition centered around 3217 $cm^{-1}$ is the main and most intense $\nu_3$ asymmetric I0> → I1> stretch, originating from both the ground vibration and orientation levels. Due to the fact that the $(L_1)_1$ excited vibration-orientation level is of *F* symmetry and that of the ground $(A_1)_1$ vibration-orientation is of *A* symmetry, the $(L_1)_1$→$(A_1)_1$ relaxation is expected to be hindered by spin freezing. The same argument applies to transitions involving the $(E_2)_1$ level which is of E nuclear spin symmetry.

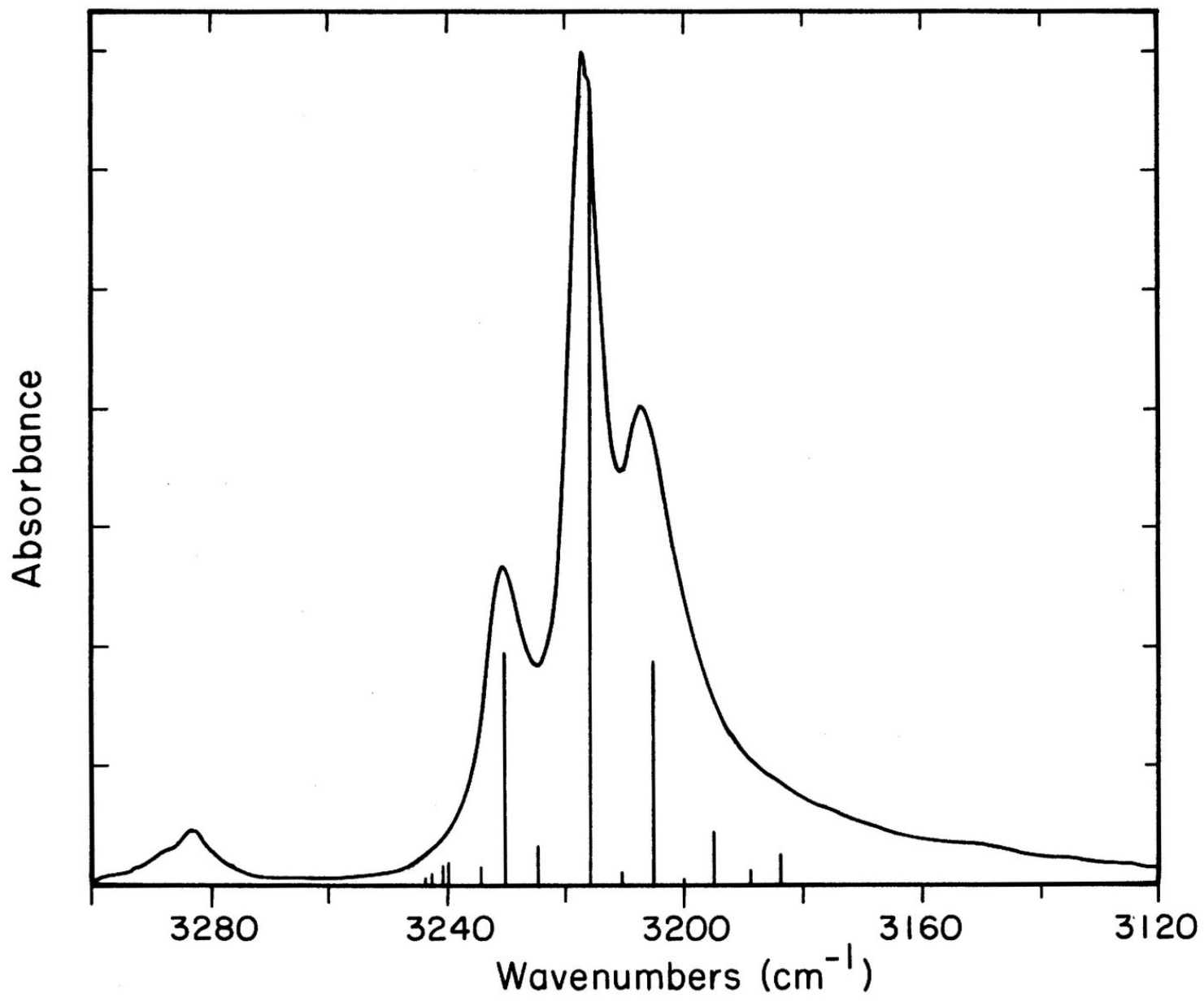

**Figure 7. The comparison between the calculated spectrum at 16 K and the experimental results at a thermometer reading of 20K. The spin temperature depends on the thermal history of the sample. The calculated intensity of the strongest band has been matched to the experimental intensity of that band. (See Ref. [8])**

Therefore, the spin freezing at these temperatures applies to the first three orientation levels of both the ground and excited vibrational states , i.e., $(A_1)_1$, $(L_1)_1$ and $(E_2)_1$. The transitions involving these levels dominate the low temperature spectra given that the first three orientation levels of the ground vibrational state are all populated at low temperatures. The slow interconversion of nuclear spins leads to the non-thermal, time-varying population distributions which explain the similarities between the spectra in Fig. 3 despite the large temperature decrease.

This spin freezing property grants the aggregation of qubits undergoing a N-H vibration-orientation outstanding uniform superposition properties given that the excited vibration-orientation states have long lifetimes. Additional evidence of the slow relaxation rate of the excited states due to spin freezing is concluded from the experimental study of the relaxation rates of the main bands at low temperatures (Ref. [8], Sec. III D).

### *c. Several Experimentally Resolvable Δω*

When two or more identical qubits are considered, an additional desired property for an ideal ensemble is to have the degeneracy of the I1> level lifted, typically by an external electric or magnetic field, so that there is a difference in frequency between the I01> and I10> transitions. In essence, the desired result is to have the I0> → I1> transition be slightly different from a qubit to its pair, so one can act as a control reference of each other. When polar symmetric tops are theoretically evaluated as candidate qubits it is proposed to lift the degeneracy of the excited state using an external electric field. The difference in frequency between the states which could be achieved if the field is applied, Δω, is of crucial importance as it the essential feature of a CNOT gate. In some cases studied, such as for the $CH_3CN$ molecule, Δω ranges from 2 to 18 kHz and only the higher values begin to be considered appropriate for qubit candidacy because of the trouble encountered in resolving single digit kHz level differences. If successful in achieving an experimentally resolvable Δω, it is proposed that entanglement could then be induced via a combination of resonant pulses which differentiate one qubit from the other in going from the I00> →I01> I10> and finally to the I11> state [18].

This $NH_4^+$ qubit should be able to fulfill the conditions to satisfy the above mentioned approach with an experimentally resolvable Δω in various ways. Consider the case that more than one frequency is selected from the envelope of the same I0> → I1> $(A_1)_1$ →$(L_1)_1$ vibration-orientation transition centered at 3217 $cm^{-1}$. The reduced case of selecting only two frequencies within that same band, recalling that they represent a large number of qubits for each frequency, would somewhat compare to the cases referenced above, as they would represent the study of a pair of identical qubit sets as a departure point. However, in principle, for the $NH_4^+$ qubit, this operation could be repeated as many times to cover the width of the band when using an excitation source narrower in frequency than the FWHM of the band, such as an IR tunable laser. This could result in various qubit-sets with an array of Δω values in the GHz region for that sole I0> → I1> $(A_1)_1$ →$(L_1)_1$ vibration-orientation transition.

A second approach is to use two distinct vibration-orientations discussed above when discussing the multiplicity of the I0> → I1> vibration-orientations. The frequency difference between two distinct peaks of the vibration-orientation bands is shown in Table I and provides a wider range of Δω values.

Additionally, there is also the possibility of using as a control transition any of the other transitions in the solid at suitable frequencies for any given experimental set-up such as the lattice modes, the lattice sum-bands to the N-H stretch, the $NH_4^+$ overtone and bending modes, any of the quantum transitions of the $NH_3D^+$ ion or any other isotopically substituted variants of the $NH_4^+$ ion or any of the transitions of the $K^+$ ion if dilute in KTPB.

### *d. Inter-qubit Distance is easily modified*

The ability to dilute in KTPB to any extent grants these qubits a simple method to regulate inter-qubit distance using this simple chemical method. The same applies to its isotopically substituted counterparts, such as $NH_3D^+$ions, which can be diluted either in KTPB or ATPB.

The multiplicity and structure of the bands in the N-H and N-D stretching is due to the individual $NH_4^+$ ions. These vibration-orientations bands remain the same upon: (i) dilution in KTPB; (ii) deuteration of the phenyl groups in ATPB to form ATPB-$d_{20}$; and (iii) on employing different crystallization techniques to prepare the samples [8]. These rule out any strong interaction between these motions and their environment and demonstrate that they are indeed due to the $NH_4^+$ ions only. Undesired inter-qubit interactions, ($NH_4^+$ - $NH_4^+$, $NH_4^+$ - $B(C_6H_5)_4^-$ and $NH_4^+$- $B(C_6D_5)_4^-$), are expected to be minimal if not absent: it is possible to rule out the possibility of any strong $NH_4^+$ - $NH_4^+$ nearest neighbors' interactions because this should have resulted in changes in the spectra upon dilution of the $NH_4^+$ ions in KTPB. For the $NH_4^+$ ion, this approach allows going from 100% ATPB to 100 % KTPB. As an example, experiments were performed with samples of 1% $NH_4^+$ in KTPB and 0.5% $NH_3D^+$ in ATPB obtaining strong signals for the quantum motions studied; these being dilution levels which in no way set limits.

The site that the $NH_4^+$ ion or the dilute $NH_3D^+$ occupies in ATPB is very large. From crystallographic data [14] taken at 120K, the average radius of the site is calculated to be over 2Å, slightly larger than the radius of an octahedral hole in a Xe matrix [8]. Distances between nearest neighboring $NH_4^+$ ions, $r_{12}$, can therefore range from about single digit, as in the case of 100% ATPB and ATPB-$d_{20}$, to hundreds of nanometers in a sample of 1% $NH_4^+$ in KTPB.

The distance $r_{12}$ becomes a key parameter in trying to achieve appropriate Δω values for the qubits being considered in the implementation of quantum logic gates for molecules trapped in an optical lattice. Typical $r_{12}$ values being considered in these theoretical studies range from single digit to hundreds of nanometers in trying to achieve appropriate levels of entanglement, coherence, dipole-dipole coupling strength for computing for each pair of qubits being considered [18-22].

The $r_{12}$ range being considered in these studies is similar to the inter-qubit distances possible in ATPB, $NH_4^+$ ions dilute to any extent in KTPB (e.g., 0,5%), and in ATPB-$d_{20}$ in both controlled (higher $D_{2d}$ character for the lattice symmetry) and uncontrolled (higher $T_d$ character for the lattice symmetry) crystallizations.

### e. Dilute $CH_4$ in a Xe Matrix

The same theoretical model used in understanding the N-H stretching $\nu_3$ vibration of the $NH_4^+$ ions in ATPB was applied to calculate the frequency and intensity of C-H stretching spectra obtained for the dilution of $CH_4$ molecules in a Xe matrix and resulted in an excellent match [15].

Therefore, if the $\nu_3$ vibration-orientation of the $NH_4^+$ ion in ATPB are considered I0> → I1> transitions of interest, the C-H stretching $\nu_3$ vibration of the dilute in a Xe matrix could be considered as alternative. They would enjoy all of the special characteristics described in this section for the N-H vibration-orientations. The main difference between the two is the magnitude of the dimensionless field-strength parameter $\beta$, which for the rotations of the $NH_4^+$ ion in ATPB had a value of 1 and in this case has a value of 5, implying a higher barrier to rotation for the $CH_4$ molecules about their $C_3$ axes [15].

Outside the C-H vibration-orientation, other motions of the solid formed by diluting $CH_4$ in a Xe matrix, including those of the Xe matrix, could result in new qubits and/or additional transitions of interest.

## 3. $NH_4^+$ rotations

As seen in Figs. 1 and 2, the low temperature Raman spectra of ATPB and ATPB-$d_{20}$ show many additional bands of typically lesser intensity between 12 and 60 $cm^{-1}$. These bands arise from the direct observation of the nearly free rotational (orientational) motion of the $NH_4^+$ ion and are assigned in Table III of Ref. [11]. These orientational transitions are the same type of quantum motion that couple to the vibration and form part of the IR N-H vibration-orientation transitions discussed previously. Fig. 3 of Ref. [11] shows that the orientational motions are present at 5K and transitions that are assigned as originating from the $(L_1)_1$ are present, thus indicating that the first excited orientation state is populated at these temperatures despite being some 12 $cm^{-1}$ above the $(A_1)_1$ state. This is interpreted as another manifestation of spin freezing, in agreement with the above analysis of the N-H vibration-orientations.
Accessing directly rotations allows for them to be considered another set of multiple I0> → I1> transitions of interest at a completely different frequency range.

# III. Summary and Conclusions

Desired pair-wise interactions such as entanglement between qubits have been achieved mainly by focusing on I0> → I1> transitions in the microwave frequency range using several techniques such as silicon spin, superconducting loops, ion traps, diamond vacancies and topological qubits. The initial goal is to incorporate a much larger number of qubits, at least 50, to be able to solve some problems classical applications cannot solve, to then aggregate hundreds and eventually thousands of qubits to do meaningful work. However, progress is slow in scaling-up the number of qubits whilst trying to achieve controlled interactions amongst them.

The approach proposed here is opposite to the ones described above. The rationale is to generate a "matrix" of qubits that traps the more mobile qubits. Therefore, the starting point is to use solids that contain an enormous number qubits from their inception. In turn, each qubit has a multiplicity of transitions of interest which can be accessed at frequencies between 10 to 3300 $cm^{-1}$ using a multiplicity of well established techniques. The more mobile qubits are molecules that exhibit at very low temperatures some of the added degrees of freedom typically associated with a molecular gas. The matrix qubits provide transitions of interest such as phonon modes and additionally, provide spatial containment for the more mobile qubits, a means to regulate their inter-qubit separation and symmetry/anisotropy to quantum motions such as vibrations and rotations.

The first scaled-up qubits being proposed using this approach comprise the components of ATPB, ATPB-$d_{20}$, KTPB and the $NH_4^+$ ion dilute in KTPB. In addition, qubit candidates which are proposed but left to be examined in further detail in future studies comprise: (i) the isotopic variants of the $NH_4^+$ ion, such as the $NH_3D^+$ ion, dilute in any of the solids mentioned above; (ii) the $K^+$ ion; and (iii) $CH_4$ either in its solid phase II or dilute in a matrix such as Xe.

As observed in the spectra in prior studies [8-12,15], each qubit has a multiplicity of I0> → I1> transitions of interest such as vibrations, rotations, overtone, bending and phonon modes.

In addition to their scaled-up nature, there are important additional advantages to using these qubits, which comprise:

1. No external fields, ultra-cold nor ultra-high vacuum conditions are required to observe the wealth of well defined and experimentally resolvable quantum transitions of these molecular solids at diverse temperatures, which in all cases is at least in the order of a few degrees Kelvin;
2. The transitions of interest have been studied both theoretically and experimentally and can be accessed at a variety of frequencies using well developed experimental techniques such as IR, Raman, NMR, neutron scattering, microwave, etc.
3. In most cases there are at least two different types of qubit per molecular solid. Each type of qubit has a multiplicity of I0> → I1> transitions of interest. Having multiple and distinct I0> → I1> transitions of interest in the solid opens up the potential to develop new ways to combine and control them;
4. It is possible to vary both the absolute number of qubits in the crystal and the inter-qubit $r_{12}$ distances from single digit to at least several hundred nanometers by chemical dilution;
5. The $\nu_3$ N-H stretching I0>→I1> transition of the $NH_4^+$ ion in ATPB, ATPB-$d_{20}$ and dilute in KTPB has several outstanding properties which when combined together with the possibility to select the inter-qubit distance make it a strong candidate to generate concurrently large entanglement, stable superposition, ease of manipulation and stability to record the results of manipulations with the available control protocols and technology in diverse experimental setups. The outstanding properties of the $\nu_3$ N-H stretching I0>→I1> transitions comprise:
    a. Several experimentally resolved frequency shifts, $\Delta\omega$, either when considering each transition in isolation or when considering the frequency difference

between vibration-orientation transitions. These are resolved for all inter-qubit distances selected;

   b. The transitions involving the first three orientation levels of the ground vibrational state, either as observed directly or in the vibration-orientation transitions, have long lifetimes and dominate the Raman and IR spectra even at very low temperatures. This is because they have different nuclear spin symmetry and the spectra imply a strong presence of nuclear spin freezing;
   c. They can be considered as several individual and distinct I0>→I1> transitions in isolation, or alternatively, as a multi-level I0>→I1>+I1'>+I1">+… vibration-orientation without the need of external fields. (This also applies to the N-D stretching I0>→I1> transition under some re-crystallization conditions (Fig. 5 of Ref. [12]); and
   d. A multi-level I0>→I1>+I1'>+I1">+… vibration-orientation should make it possible to carry diverse quantum information signals at once. This experimentally verified multilevel transition should make it is possible to conceive combined states beyond the pairwise I00> → I11> transitions and logic protocols that go beyond the binary system. It should also allow applying full quantum cryptography.
6. $CH_4$ either in its solid phase II or dilute in a Xe matrix is also proposed as a qubit. The $\nu_3$ C-H stretching vibration-orientation in these solids is analogous to the $\nu_3$ N-H stretching I0>→I1> transition of the $NH_4^+$ ion discussed above [15]. Outside the C-H vibration-orientation, other motions of the solid formed by diluting $CH_4$ in a Xe matrix, including those of the Xe matrix, could result in new qubits and additional transitions of interest;
7. The tetraphenylborate lattice has well defined quantum motions at temperatures ranging from single digit to 300 K which are distinct to the ones performed by $NH_4^+$ion. This is an exceptional temperature range for transitions of this kind and may make it very attractive to several applications. The $B(C_6H_5)_4^-$ and $B(C_6D_5)_4^-$ anions are proposed as qubits on their own;
8. Solids grant more mobile qubits spatial order within a given crystal. Additionally, solids can also grant qubits distinct locations in space when forming an array of crystals. The possibility to differentiate otherwise identical transitions in space opens the possibility, amongst other things, to develop novel hardware and control gates, parallel computing, split photon excitations to generate entanglement of molecular transitions at a distance. The application of these qubit candidates to teleportation is mentioned here and dealt in more detail in a future paper [23];
9. All qubits have hydrogen atoms as components which allows for great sensitivity in neutron scattering and to NMR pulses and control techniques. Additionally, the possibility to partially or totally deuterate the qubit can be utilized to alter the multiplicity and frequency of the transitions of interest; and
10. The $K^+$ ions in KTPB are available as another type of qubit. It would represent the example of trapping an atomic ion, rather than a molecular ion, in the tetraphenylborate matrix. Its specific transitions of interest are not studied here and hence their properties are additional to the above considerations.

The various qubits presented in this study should require no further significant development to make diverse applications a reality in the short term.